\documentstyle[12pt]{article}

\def\a{\begin{eqnarray}}
\def\b{\end{eqnarray}}
\def\0{\nonumber}
\input amssym.def
\font\teneusm=eusm10                    
\font\seveneusm=eusm7                   
\font\fiveeusm=eusm5                    
\newfam\eusmfam
\textfont\eusmfam=\teneusm
\scriptfont\eusmfam=\seveneusm
\scriptscriptfont\eusmfam=\fiveeusm

\input amssym
\newcommand{\be}{\begin{equation}}
\newcommand{\ee}{\end{equation}}
\newcommand{\ba}{\begin{eqnarray}}
\newcommand{\ea}{\end{eqnarray}}
\newcommand{\ban}{\begin{eqnarray*}}
\newcommand{\ean}{\end{eqnarray*}}
\newcommand{\brr}{\begin{array}}
\newcommand{\err}{\end{array}}
\newcommand{\bc}{\begin{center}}
\newcommand{\ec}{\end{center}}

\newcommand{\pa}{\partial}
\def\A{{\cal A}}

\def\E{{\cal E}}

\def\H{{\cal H}}
\def\L{{\cal L}}
\def\La{\Lambda}
\def\M{{\cal M}}

\def\R{{\cal R}}

\def\W{{\cal W}}

\def\l{\lambda}

\def\al{\alpha}

\newcommand{\bea}{\begin{eqnarray}}
\newcommand{\eea}{\end{eqnarray}}
\newcommand{\bean}{\begin{eqnarray*}}
\newcommand{\eean}{\end{eqnarray*}}
\newcommand{\F}{{\cal F}}

\begin{document}

\begin{titlepage}

\begin{flushright}
SISSA-ISAS 32/93/EP
\end{flushright}
\vskip0.5cm
\centerline{\LARGE  W--Infinity Structure}
\centerline{\LARGE of the $sl(N)$ Conformal Affine Toda Theories}
\vskip1.5cm
\centerline{\large  R. Paunov}
\centerline{International School for Advanced Studies (SISSA/ISAS)}
\centerline{Via Beirut 2, 34014 Trieste, Italy}
\vskip5cm
\abstract{We reexamine the $W_{\infty}$ symmetry of the $sl(N)$
Conformal Affine Toda  theories. It is shown
that it is possible to reduce (nonuniquely) the zero curvature equation
to a Lax equation for a first order pseudodifferential oprator,
whose coefficients are the generators of the $W_{\infty}$
algebra.
This clarifies the known relation between
the Conformal Affine Toda  theories and
the KP hierarchy. A possible correspondence between the
matrix models and the Conformal Affine Toda models is
discussed.}

\end{titlepage}


\section{Introduction}

The study of the higher--spin extensions of the Virasoro algebra
has acquired a central role in the two--dimensional physics.
In the conformal field theory \cite{Za}  the $W_{n}$ algebras
appear as  chiral algebras of a huge set of rational
conformal field theories. In their classical version the
$W_{n}$ algebras underlie the integrability structure of
the $n$th Korteweg--de Vries (KdV)--type hierarchy and related to them matrix
Drinfeld--Sokolov (DS) hierarchies \cite{DS},\cite{Di}.
Both of these two types of
hierarchies are integrable hamiltonian systems possesing a pair
of Poisson  brackets  which are coordinated (i. e. any linear
combination of them is again a Poisson bracket).
The  integrable hierarchies of KdV--type arise
in the scaling limit of the matrix models of the
two--dimensional gravity coupled to $c\leq 1$ conformal matter \cite{BK}
(for a review see \cite{Lag}) and have a deep geometrical interpretation
\cite{Wi}. Recently it has been realized
that it is not nesessary to go to the scaling limit
in order to recover integrable hierarchies \cite{BX}.

A natural generalization of the finite classical $W$ algebras appears
when one takes the limit $n\rightarrow \infty$. In this limit the
nonlinear part of the second Gelfand--Dickey (GD) bracket disappears
and  one obtains the so called $w_{\infty}$ algebra \cite{Ba}. Another
way to construct $W$--infinity algebras is to pass directly to the infinite
generalization of the KdV--type hierarchies, the Kadomtsev--Petiashvili (KP)
hierarchy. The first GD bracket \cite{Wa}  produces the $W_{\infty}$ algebra
studied in \cite{Po} . The nonlinear generalizations  of
$W$--infinity algebra  are related to the second GD bracket.
In \cite{AFGZ} a large class of $W$-infinity algebras are
realized in terms of two fields of spin one and two. This
construction applies to the Conformal Affine Toda (CAT) theories \cite{BB}.
The $W$--symmetry of the CAT models is a strong indication that
they are integrable and that their integrability structure is
dictated by the KP hierarchy.

In this paper  the $W_{\infty}$ symmetry of  $sl(N)$
CAT is reconsidered from the point of view of the inverse scattering method.
Starting from the spatial component of the Lax connection  we show that
after a suitable gauge transformation it assumes a very simple
form: its matrix part belongs to the Heisenberg subalgebra
associated with the principal gradation of the affine Lie algebra
$\hat{sl}(N)$ plus a term proportional to the derivation of $\hat{sl}(N)$.
Our approach is similar to the approach used in
\cite{Hol}, \cite{HM}  to construct a large class of DS--type
integrable hierarchies. The main result in these papers is that for
each positive graded element of a Heisenberg subalgebra of an untwisted
affine Lie algebra there exist a collection of hierarchies related through
Miura maps. Therefore,  modulo Miura transformations, the inequivalent
hierarchies are given by  inequivalent Heisenberg subalgebras
(the classification of the graded regular elements which belong to certain
Heisenberg subalgebra of   the loop algebra $\tilde{gl}(n)$
has been done in
\cite{FeHM}). The situation in CAT is however different.
In this way one can produce only a finite nuber of conserved
quantities. A further reduction in getting the $W$--infinity
currents consists in restricting the Lax connection on the
Fock space associated with grade one element (and its conjugate)
of the Heisenberg subalgebra. On this subspace the matrix part of the
Lax connection is closely related to the spectral matrix
in the one--matrix models \cite {BX}. Therefore
it seems reasonable to state that the multiplication by
the spectral parameter in the
matrix models can be identified with the differentiation along the
space direction in CAT.
The problem of the reduction to the  KP hierarchy is also
considered. It turns out that this reduction is not unique.
On the other hand, it has been proven recently \cite{ANPV}
that both the two-- and the four--boson KP hierarchies are gauge
equivalent to the standard KP hierarchy.

\section{The Drinfeld-Sokolov Reduction}

This section is devoted to a review of the  results
of \cite{DS}. The basic idea is very simple: a system of
$n$ first--order differential equations $\L\xi=0$
($\xi$ is a $n$ dimensional vector) for a certain class
of first--order matrix  differential operators can be reduced
to a $n$th order {\it scalar} differential equation
$L\xi=0$.  Drienfeld and Sokolov considered
the case when $\L$ has the following form
\a
\L=\pa_x+Q(x)+\La
\label{auxilary}
\b
where $Q(x)$ is a {\it lower} triangular $n\times n$ matrix  and
$\La=\E_++\l \E_-^{n-1}$, $\La^{-1}=\E_-+\l^{-1} \E_+^{n-1}$.
Here and in what follows we shall use the notations
$\E_+=\sum_{i=1}^{n-1} E_{i i+1}$, $\E_-=\sum_{i=1}^{n-1} E_{i+1 i}$
and $E_{ij}$ is the matrix having one at $(i,j)$-th site
and zero elsewhere.
The extra variable $\l$ is usually called the spectral
parameter; $Q(x)+\La$ will be called the matrix part of $\L$.
One easily checks that the identities
$\La^{\pm i}=\E_{\pm}^i+\l^{\pm 1} \E_{\mp}^{n-i}$
for $i=1,\ldots n-1$  and  $\La^n=\l$  are valid.
The adjoint action of the element $\La$ permutes
the diagonal matrices
$\La E_{ii} \La^{-1} = E_{i-1 i-1}$.

The evolution of $\L$ is determined by  the Lax equation
\a
\pa_t \L =[\A , \L]
\label{Lax}
\b
which implies that the spectrum of $\L$ does not depend
on the time parameter $t$. The operator $\A$ in the Lax
equation cannot be arbitrary. In order to get a consistent
time flow    one should require that
the commutator $[\A , \L]$ has  the same form as
the matrix $Q$.

There is a standard procedure (see for example \cite{DS},
\cite{Di}) to build  up evolution equations
for the operator $\L$.  Given a resolvent $\M$ of
$\L$, i. e. a formal Laurent expansion
$\M=\sum_{i=r}^{\infty} \M_i \La^i$
with coefficients being diagonal matrices\footnote{It is easy to
show that any Laurent polynomial on $\l$ can be uniquely
rewritten in powers of $\La$ with diagonal coefficients}
which commutes with $\L$,
one sets $\A=\M_+=\sum_{i\geq 0} \M_i \La^i$ (the negative
part of $\M$ is defined as $\M_-= \sum_{i\leq -1} \M_i \La^i$).
For generic lower diagonal $Q$ this choice produces
a consistent flow since $[\M_+,\L]$ is a polynomial on $\l$
while $[\M_-,\L]$ contains only nonpositive powers of $\l$
and its free term results to be a lower diagonal matrix.
In order to find the space of the resolvents $\R_{\L}$
of $\L$ consider the gauge transformation
\a
\L^G =G  \L  G^{-1}=\pa_x G G^{-1} + G \L G^{-1}
\label{gauge}
\b
such that
\a
\L^G=\pa_x + \La + \sum_{i=0}^{\infty} h_i(x) \La^{-i}
\label{canonical}
\b
where $h_i$ are {\it functions}\footnote{This approach is
alternative to the direct diagonalization of the matrix part
of $\L$ considered in \cite{FT}}. The  solution is
given by  the expansion
\a
G(x,\l)=1+\sum_{i=1}^{\infty} g_i(x) \La^{-i}
\b
where $g_i$ should satify the recursion relations
\a
h_i(x) +\La g_{i+1}(x) \La^{-1}-g_{i+1}(x)=
Q_i(x)+\sum_{p=0}^{i-1}g_{i-p}\La^{p-i} Q_i \La^{i-p}-\0\\
-\frac{\pa g_i(x) }{ \pa x}
-\sum_{p=0}^{i-1} h_p(x)\La^{-p} g_{i-p}\La^p
\label{reccursive}
\b
Any diagonal matrix $S$ can be represented (nonuniquely)
as $S=Tr S + \La g \La^{-1} -g$
where $g$ is diagonal and therefore
there exists a (formal) gauge transformation
which brings $\L$ into the form (\ref{canonical}). The advantage
gained by this gauge transformation is that the resolvent
$\R_{\L^G}$ has a very simple form: it consists of all
the expansions $\sum_{i} m_i \La^{i}$
where the  coefficients $m_i$ are constant scalar matrices.
Therefore the operators
\a
\A=\sum_{i\geq 0}m_i \left( G \La^i G^{-1} \right)_+
\label{flusso}
\b
provide consistent equations of motion (\ref{Lax}).
Moreover, from the gauge transformed Lax equation
$\pa_t \L^G=[\A^G , \L^G]$  where
$\A^G=\pa_t G G^{-1} +G \A G^{-1}=\sum_i a_{-i}^G(x) \La^i$ it follows that
all the coefficients in the expansion of $\A^G$
in powers of $\La$ are scalar matrices. The coefficients
$a^G_{-i}$ are constants on $x$ for negative $i$
and $\pa_t h_i+\pa_x a^G_{-i}=0$ for $i\geq 0$. Therefore
the evolution equation (\ref{Lax}) possesses an infinite number
of conserved quantities
\a
I_l=\int h_l(x) dx
\label{cariche}
\b
Another intriguing property  of the flows generated by
the operators (\ref{flusso}) is that they commute. More precisely,
if $\M_1$ and $\M_2$ are resolvents of $\L$, $\A_i$ are
their positive parts $\A_i=\left( \M_i\right)_+$, then
the evolution equiotions $\pa_i L=\pa_{t_i} L=[ (\A_i)_+ ,\L]$
are in involution $[\pa_i,\pa_j]L=0$. The last identity
is equivalent to the condition that the curvature
$F_{12}=\pa_1 \A_2-\pa_2 \A_1 -[\A_1,\A_2]$ is a resolvent of
$\L$. A stronger result is valid, this curvature vanishes
identically. In proving this one
first observes that $\pa_i \M_j=[\A_i,\M_j]$  and
therefore $F_{12}=[\A_1,\M_2]_++[\M_1,\A_2]_++
[\A_1,\A_2]$ which vanishes since $\M_1$ and $\M_2$
are commuting.

Techincally it is quite cumbersome to work with
general lower triangular matrices $Q$ in (\ref{auxilary}).
There exists a gauge transformation   $S$ such that the
gauge transformed matrix $U=Q^S=\pa_x S S^{-1}+
S( Q+\La) S^{-1} -\La$ has only one non--vanishing row:
the last one \cite{DS}
\a
U=-\sum_{k=0}^{n-1} u_{n-k}E_{n-1\, k}
\label{vudoppio}
\b
Till the end of this section we shall skip the index
$S$ for brevity $\L=\pa_x +\La + U$.
It is instructive to return to the problem of the
construction of consistent evolution equations in
this gauge. First we introduce some useful notations.
It will be convenient \cite{Di} to consider matrices
as operators in the space of integral (Voltera) operators
spanned by the symbols $\pa^{-i}$ ($i\geq 1$) factorized
by the subspace of pseudodifferential operators (PDO)
of degree less than $-n$. The multiplication is defined by
the generalized Leibniz rules
\a
\pa^{-i-1} f= \sum_{l=0}^{\infty} (-)^{l}
\left(\brr{c}i+l \\ l \err\right) f^{(l)}\pa^{-i-l-1} ~~~~~~~~~~
f^{(l)}=\frac{{\pa}^l}{\pa x^l}f\0
\b
for arbitrary function $f$ on $x$. Given a matrix with entries $A_{ij}$,
$i,j=0,1,\ldots n-1$  the vector--columns  and vector--rows  are
introduced as follows
\a
A^i=\sum_{l=0}^{n-1} \left(-\pa \right)^{-1-l}A_{li}~~~~~~~~~~~
A_i= \sum_{l=0}^{n-1}A_{il} \left(-\pa \right)^l
\b
In these notations the multiplication of two matrices can be
written as $(AB)^j=\sum_i (-\pa )^{-i-1} res(A_iB^j)$,
$(AB)_i=\sum_j  res(A_iB^j) (-\pa )^j$ where the residue
of the PDO $X=\sum X_i (-\pa )^i$ is defined to be $X_{-1}$.
The consistensy of (\ref{Lax}) implies $[\L, \A]_i=0$,
$i=0,\ldots,n-2$  and the equations of motions read
$\pa_t U_{n-1}=-[\L, \A]_{n-1}$. These equations can be written
in the following form
\a
\A_{i+1}=-\pa \A_i -\A_{in-1} L(\l)&&~~~~~~~~
\A_n=\pa_t U_{n-1}-\sum_{l=0}^{n-1} u_{n-l}(\l)\A_l\0\\
&&\hskip -3cm L(\l)= \left(-\pa \right)^n+\sum_{i=0}^{n-1} u_{n-i}(\l)
\left(-\pa \right)^i
\label{rec}
\b
where $u_{n-i}(\l)=u_{n-i}-\l \delta_{i,0}$.
The solution of the upper recursion relation is
\a
\A_i=\left( -\pa \right)^i \A_0-\left( (-\pa)^i \A^{n-1} \right)_+L(\l)
\label{sol}
\b
and $X_+$ is the differential part of a PDO $X$; $X_-=X-X_+$ is
expanded on negative powers of $\pa$.
Setting $i=n$  in (\ref{sol}) and taking into account that due to
the second eq. (\ref{rec}), $\A_n$
is a differential operator of order not not greater than $n-1$
one gets $\A_0=\left( \A^{n-1} L(\l) \right)_+$ and therefore the
equation of motion can be written in the form
\a
\frac{\pa L}{\pa t}&=& - H(\l)(\A^{n-1} )~~~~~~~~
H(\l)(X)=\left( L(\l) X\right)_+ L(\l)-L(\l) \left( X L(\l) \right)_+
\label{Adler}
\b
$H(\l)$ is the Adler map\footnote{due to the identity
$H(\l)(X)=-\left( L(\l) X\right)_- L(\l)+L(\l) \left( X L(\l) \right)_-$
it follows that the immage of the Adler map belongs to the set of
the differetial operators of order not greater than $n-1$. The
Adler map annihilates the pure differential operators and the PDO
of order less than $-n$. }. This completes the reduction of the
matrix evolution equation (\ref{Lax}) to an evolution equation
for a $n$th order (scalar) differential operator.
Note also that the Adler map splits into two terms $H(\l)=H_2+\l H_1$
where:
\a
H_1(X)=[X_-,L]~~~~~~~~~
H_2(X)=\left( L X\right)_+ L-L \left( X L \right)_+\0\\
\b
where $L=L(0)$ and the following recursion relations are obviuosly valid
\a
H^0(L^{\frac{r-n}{n}})+H^{\infty}(L^{\frac{r}{n}})=0
\label{biham}
\b
For $\l \rightarrow \infty$ and $\A^{n-1}=L^{\frac{m}{n}}$ in
(\ref{Adler}) the equations of motion coincide with the $m$th
flow of the $n$th KdV--type hierarchy:
\a
\frac{\pa L}{\pa t_m}&=&-[L^{\frac{m}{n}}_-,L]=[L^{\frac{m}{n}}_+,L]\0\\
\label{KdV}
\b
which are hamiltonian with respect to the Gelfand--Dickey (GD)
brackets
\a
\{<L,X> , <L,Y> \}_i^{GD}=<H_i(X) Y>~~~~~~~
<L,X>=\int res LX dx
\label{GD}
\b
The hamiltonans $I_{m+n}=-\frac{n}{m+n}\int res L^{\frac{m+n}{n}}$
generate the $m$th KdV flow with respect to the first GD bracket
while the corresponding hamiltonian with respect to the second GD
bracket is $-I_m$. This  is seen from the recursion relations
(\ref{biham})  written in terms of the GD brackets
\a
\{I_{i+n}, L\}_1+\{I_i,L\}_2=0
\b
Using the recursion relations it is easy to show that the hamiltonians
$I_i$ are in involution with respect to the both GD brackets.  On the
other hand the equations (\ref{rec}) allow from a given PDO $X$ to
reconstruct the matrix $\A(X)$ (using the identification between
vectors and PDO's the coefficients of $X$  are related to the
elements of the $n$th column of $\A$ through the relation
$\A^{n-1}=X$). Moreover, the Drienfel-Sokolov (DS) brackets coincide with the
GD brackets.
\a
\{ <U,\A(X)>, <U, \A(Y)> \}^{DS}&=&\int Tr [\L, A(X)] \A(Y) dx=\0\\
=\int res H(\l)(X) Y dx&=&\{<L,X> , <L,Y> \}^{GD}
\label{DS--GD}
\b

We shall leave this section with the following remark.
The coefficients $h_i$ in the expansion
(\ref{canonical}) and the KdV hamiltonian densities  $res L^{\frac{i}{N}}$
are not independent. For each $i=1,2,\ldots$ there
is a linear combination of these two densities which is a total derivative
 ( see \cite{DS} for a
detailed proof).

\section{Generalization to the $sl(N)$ Conformal Affine Toda Theories}

In this section we shall generalize the approach from the
previous section to the $sl(N)$ Conformal Affine Toda theories
\cite{BB}. The equations of
motion can be written in a Lax form
(\ref{Lax}) but the matrix part of $\L$
belongs to  the affine  Lie algebra $\hat{g}=\hat{sl}(N)$.
Therefore we shall need some Lie algebraic background.
As a basis in $\hat{g}$ we choose the generators $E^{ij}_k$
 ($i,j=1\ldots N, \,\,\, i\neq j$), $H^{\al_i}_k, \,(\,\, i=1,\ldots N-1=
rank sl(N))$, $\hat{c}$ and $\hat{d}$ with the
commutation relations
\a
[E^{ij}_n,E^{kl}_m]=\delta^{jk}E^{il}_{n+m}-\delta^{li}E^{kj}_{n+m} +
n\hat{c} \delta^{jk}\delta^{il}\delta_{n+m,0}\0\\
\b
\a
\hskip -0.5cm [ H^{\al_i}_n,E^{kl}_m ]=\left(\delta^{ik}-\delta^{i+1k}-
\delta^{il}+\delta^{i+1l}\right)E^{kl}_{n+m}\0 ~~~~
[\hat{d},E^{ij}_n]=\left( Nn +j-i\right) E^{ij}_n\0
\b
\a
\hskip -2cm [ H^{\al_i}_n, H^{\al_j}_m ]=n\hat{c}\left(2 \delta^{ij}-
\delta^{ij+1}-\delta^{i+1j}\right)~~~~~~~~~~~~~~~~
[\hat{d},H^{\al_i}_n]=Nn H^{\al_i}_n
\label{KacMoody}
\b
and $\hat{c}$ is the central element.  The derivation $\hat{d}$
corresponds to the {\it principal} gradation \cite{Kac}.
We choose the invariant inner product as follows
\a
<E_n^{ij} , E_m^{kl} >&=&\delta_{n+m,0} \delta^{jk} \delta^{il}
{}~~~~~~~~~ < H^{\al_i}_n , H^{\al_j}_m > = \delta_{n+m,0} K_{ij}\0\\
< \hat{d} , \hat{c} >&=& N ~~~~~~ < \hat{d} , H_0^{\al_i} >= 1
{}~~~~~~~< \hat{d} , \hat {d} >= \sum_{i,j=1}^{N-1} K^{ij}
\label{inner}
\b
where $K_{ij}$ and $K^{ij}$ are the entries of the Cartan matrix
and of its inverse. The gradation in $\hat{g}$ is introduced naturally
through the adjoint action of $\hat{d}$
\a
\hat{g}=\oplus_{i\in Z\!\!\!\!Z} \hat{g}_i~~~~~~~~~~
[\hat{d},\hat{g}_i]=i\hat{g}_i
\label{grado}
\b
The elements
\a
\E_{nN+i}=\sum_{r=1}^{N-i}E^{r \, r+i}_n+\sum_{r=1}^{i}
E^{N-i+r\, r}_{n+1}~~~~
\E_{-nN-i}&=&\sum_{r=1}^{N-i}E^{r+i\,r}_{-n}+
\sum_{r=1}^iE_{-n-1}^{r\, N-i+r} \0
\b
of grade $\pm (Nn+i)$ ($i=1,\ldots N-1$) and
$\hat{c}$ generate the  Heisenberg subalgebra
\a
[\E_n,\E_m]=n\hat{c}\delta_{n+m,0}
\label{Heisenberg}
\b
The gradation defined by $\hat{d}$ is associated to this
Heisenberg subalgebra in the sence that $\E_i\in \hat{g}_i$.

The subspaceses $\hat{g}_{\pm(Nn+i)}$ are spanned by the
elements ${E^{r\, r\pm i}_{\pm n}}$ and
${E^{\pm N \mp i+r\, r}_{\pm(n+1)}}$ for
$i=1,2,\ldots , N-1$ and $n=0,1,\ldots$ ;
$\hat{g}_{\pm Nn}$    are spanned by $H^{\al_i}_{\pm n}$
for $n$ being positive integer number; the zero grade
subspace consists of the elements $H^{\al_i}_0$, $\hat{c}$ and
$\hat{d}$. Moreover
\a
\hat{g}=Ker(ad(\E_1))\oplus Im(ad(\E_1)) \oplus \hat{d}
\label{regulare}
\b
This property will be important in what follows.

The Conformal Affine Toda (CAT) models (\cite{BB}) arise as a
conformally invariant extension of the affine Toda theories.
The equations of motion are equivalent to the flatness of
the connection
\a
\L_+=\pa_++2\pa_+\Phi +\E_1~~~~~~~
\L_-=\pa_-+e^{-2ad \Phi} \E_{-1}\0\\
\b
where $\Phi= \frac{1}{2}\sum_{i=1}^{N-1} \phi_i H^{\alpha_i}_0
+\eta \hat{d} +\frac{1}{2} \xi \hat{c}$. Here and in what follows
$x^{\pm}=x\pm t$.
In tems of the components of the field $\Phi$  one gets
\a
\pa_+\pa_- \Phi_a&=& e^{\sum_{a=1}^{N-1} K_{ab}\phi_b+2\eta}-
e^{-\sum_{b=1}^{N-1} K_{\theta b} \phi_b + 2\eta }\0\\
\pa_+ \pa_- \eta&=&0\0\\
\pa_+ \pa_- \xi &=& e^{-\sum_{b=1}^{N-1} K_{\theta b} \phi_b + 2\eta }
\label{CAT}
\b
where $K_{\theta b}=2 \frac{ \theta \cdot \al_b}{\theta \cdot \theta}$,
$\al_b$ are the simple roots and $\theta$ is the highest root
( for $sl(N)$ $\theta=\al_1 +\ldots +\al_{N-1}$).

In the previous section we learned that from a  given  Lax operator
$\L=\L_x=\L_++\L_-$ (\ref{auxilary}) one can construct infinite set
of integrals of motion. The strategy was to  perform a gauge transformation
such that the gauge transformed connection $\L^G$ belongs to a
maximally commutative subalgebra $\H$  of the loop algebra $\tilde{gl}(n)$.
In the case considered in Sec.2 this algebra is generated by
the integer powers of the matrix $\La$. An important point in this approach is
that the highest grade term  $\La$ of $\L$ is a {\it regular}
element, i. e. the   loop algebra  decomposes into a direct
sum of $Ker(ad \La )=\H$ and $Im (ad \La)$. This allowed
to show that there exists a solution of the recursion relations
(\ref{reccursive}). The situation in CAT looks to be quite similar
sinse $\L_x$ has the same form as (\ref{auxilary}) but the  presence
of the central element and the derivation $\hat{d}$ have to be
taken into account. The case when $\L$ does not contain
$\hat{d}$ is considered in \cite{HM}. One could expect that after
adding such a term the situation will drastically change. The
reason is that $\hat{d}$ is not in the immage of the adjoint
action of the highest grade element in $\L$ (in our case $\E_1$).
It is therefore natural to look for a gauge transformation such that
the gauge transformed connection has the form
\a
\L_x^G&=&\E_1 +\pa_x +J(x) \hat{d}+
\sum_{\stackrel{i=1}{i\neq 0(mod N)}}h_i(x) \E_{-i}
\label{Lx}
\b
where $G=e^{\sum_{i=1}^{\infty} g_{-i}}$, $g_{-i}\in \hat{g}_{-i}$
and $J(x)=2\pa_+ \eta$. For $g_{-i}$ and $h_i$ we obtain the following
recursion relations
\a
 J(x) \hat{d}+ [ \E_1 , g_{-1} ]  &=& 2 \pa_+ \Phi
\label{primo}
\b
\a
h_l \E_{-l} + [ \E_1 , g_{-l-1}] &=&
P_l ~~~~l\geq 1
\label{generale}
\b
where  $P_l $ depends on $g_{-1},\ldots , g_{-l}$
(for $l=0(mod N)$ the first term in (\ref{generale}) is
missing).
These recursion relations are solvable due to (\ref{regulare}).
It is easy to get
\a
g_{-1}&=&-\sum_{i=1}^{N-1} \pa_+ (\phi_i +\xi) E_0^{i+1i}-
\pa_+ \xi E^{1N}_{-1}\0\\
<\E_1,\E_{-1}> h_1&=& \frac{1}{2}
<2\pa_+\Phi + J\hat{d},2\pa_+\Phi - J\hat{d}>
-\frac{\pa}{\pa x} < g_{-1},\E_{1}>+\0\\
&&+<e^{-2ad\Phi} \E_{-1},\E_1>
\b
 From the equations of motion (\ref{CAT}), (\ref{primo}) and
(\ref{generale}) it follows that
\a
\L_t^G=G\L_tG^{-1}=\E_1+\pa_t +J(x) \hat{d} + h_1 \E_{-1} +
\sum_{i\geq 2 } a_i ~~~~~~~~~~~~~
a_i \in \hat{g}_{-i}
\label{Lt}
\b
Substituing in the
commutator  $[ \L_x^G , \L_t^G ]=0$ the expansions
(\ref{Lx}) and (\ref{Lt}) we
see that it vanishes if $\pa_-J=\pa_-h_1=0$ and
all $a_i$ in (\ref{Lt}) belong to the Heisenberg
subalgebra but  $h_i$ for $i\geq 2$ are {\it not}
conserved.

In order to built up the rest of the conserved currents
we have to use a different approach.
Let $v$ be a lowest weight vector of $\hat{g}$, i. e.
$v$ is annihilated by all the negative degree elements
of $\hat{g}$ ( in particular, by $\E_{-i}$, $i\geq 1$).
Denote by  $\F$ the Fock space generated by the
action of $\E_1$ on $v$. It is obvious from (\ref{Lx}) and
our conclusion that $a_i$ in (\ref{Lt}) belong to the Heisenberg
subalgebra that $\F$ is invariant under the action of
$\L^G_x$ and $\L_t^G$ and that on $\F$
\a
\L_x^G=\L_t^G
\label{chiral}
\b
Moreover, any element of $\F$  can be uniquely
expressed as a polynomial on $\L_x$ acting on
$v$
\a
\E_1v&=& (\L_x^G-d_0)v ~~~~~~~~~~~~~ \hat{d}v=d_0v \0\\
\E_1^nv&=&\left( (\L_x^G)^n + \sum_{k=1}^n
u_k(n) (\L_x^G)^{n-k}   \right) v
\label{charges}
\b
where the functions $u_k(n)$ satisfy the following recursion
relations
\a
u_k(n+1)=u_k(n)+\left( \pa_x -(n+d_0)J(x) \right)
u_{k-1}(n) + cnh(x)u_{k-2}(n-1) \0
\b
where $\hat{c}\F=c\F$;  $h=h_1$.
 From the Lax equation on the Fock space $\F$ it follows that
$\pa_t u_k(n)= \pa_x u_k(n)$ and therefore $u_k(n)$ are
densities of conserved currents.

We first present a heuristic derivation of $W_{\infty}$
currents. We shall further set $c=1$.
Denote by $V$ a matrix solution of the auxilary linear
problem $\pa_x V= V(\E_1 + J(x) \hat{d}+\E_{-1})$. From
(\ref{Lx}) one obtains the equations
\a
\left(\frac{\pa}{\pa_x} - d_0 J(x) \right) V(x) v&=&V(x) \E_1 v\0\\
\left(\frac{\pa}{\pa x} - (d_0+1) J(x) \right)\E_1 V(x)+
h(x)V(x) v &=& V(x)\E_1^2v
\b
and therefore
\a
LV(x)\E_1v=V(x)\E_1^2v~~~~~~L=\pa_x -(d_0+1)J(x)+
h(x)\frac{1}{\pa_x-d_0J(x)}
\label{pseudo1}
\b
Using the generalized Leibniz rule we expand
the  first order PDO  $L$ in powers of $\pa_x$
\a
L=\pa_x -(d_0+1) J(x)-
\sum_{s=1}^{\infty}  u_{s-1} (x) (-\pa_x)^{-s}
\b
The coefficients $u_s$ are expressed in terms of the
Fa\'{a} di Bruno polynomials
$P_i(J)= (\pa_x +  J(x))^i(1)$
\a
u_i(x)=h(x)P_i(-d_0J)  ~~~~~~~ i=0,1,\ldots
\label{vuinf}
\b
In this form the generators of the $W_{1+\infty}$ algebra appeared
in \cite{AFGZ}. It is worthwhile to note that
 nevertheless that the KP $L$--operator appears
in the left hand side of (\ref{pseudo1}), this equation is
{\it not} a scalar pseudodifferential equation.  Therefore it seems that
the KP structure in CAT is  encoded in a  different manner.

In trying to construct the KP hierarchy within CAT we
first note that after a suitable
$GL(\infty)$\footnote{ we define $GL(\infty)$ as the
group of all semi--infinite invertible matrices}
gauge transformation
the connection (\ref{Lx}) assumes the form
\a
\L_U=\pa_x +\E_1+U~~~~ U=-\sum_{i=1}^{\infty} u_{i-1} E_{0 i}
\b
which is a natural generalization of (\ref{vudoppio}).
The generators $\E_{\pm 1}$ and $\hat{d}$ are expressed in terms of
the elementary semi--infinite matrices as follows
\a
\E_1=\sum_{i=0}^{\infty}  E_{i+1i}~~~~~~~
\E_{-1}=-\sum_{i=0}^{\infty}(i+1) E_{ii+1}~~~~~~
\hat{d}=\sum_{i=0}^{\infty}iE_{ii}+d_0
\label{semi}
\b
The difference with respect to the finite dimensional case is that neither
the gauge transformation which brings (\ref{Lx}) into the upper form
nor the functions $u_i$ are uniquely fixed. Instead of
(\ref{vuinf}) one could take $\tilde{u}_i=u_i+P_i$,
where $P_i$ are spin $i$ differential polynomials on $u_{i-1},\ldots
u_{0},J$. This problem has been discussed in \cite{Ar} from the
point of view of the hamiltonian reduction of the two--loop
WZNW models.
Similarly to the previous section we introduce the vector--columns and
vector--rows
\a
A^i=\sum_{l=0}^{\infty} \left(-\pa \right)^{1-l}A_{li}~~~~~~~~~~~
A_i= \sum_{l=0}^{\infty}A_{il} \left(-\pa \right)^{-l}
\b
Looking for a consistent time evolution equations
$\pa_t U=[\A, \L]$ with given $\A^0$ together with the
initial condidition $\A_0=\A^0_-L-(\A^0L)_-$ where
$L$ is the PDO
\a
L=-\left(\pa\right) +\sum_{i=0}^{\infty}u_{i-1}\left(-\pa\right)^{-i}
\b
we get the solution
\a
\A_i=\left(-\pa\right)^{-i} \left(A^0L\right)_+-
\left(\left(-\pa\right)^{-i}A^0\right)_+L
\b
and the equations of motion can be written in the form (\ref{Adler})
\a
\frac{\pa L}{\pa t} = -H_2(\A^0) ~~~~~
H_2=\left(LX\right)_+L-\left(LX\right)_+L
\label{KP2}
\b
Setting in the upper equation $\A^0=L^{k-1}$ one recognises
(up to a sign) the  $k$th flow of the KP
hierarchy. The same flow can be also reproduced from the
Adler map $H_1(X)= [L,X_+]-[L,X]_+$. This follows from the recursion
relations  $H_1(L^k)+H_2(L^{k-1})=0$. $H_1$ and $H_2$ define
the first and the second GD brackets
\a
\{ <L, X>, <L,Y>\}^{GD}_i=<H_i(X),Y>~~~~~~~~<X,Y>=\int res XY dx
\b
The KP equations are hamiltonian with respect to these Poisoon
brackets. The corresponding hamiltonians are proportional
to $\int res L^k dx$. Their involution follows from the
recursion relations for the two Adler maps $H_1$ and $H_2$.
The first KP--flow implies that the coefficient of the
KP operator are chiral $\pa_t u_i=\pa_x u_i$.
Similarly to (\ref{DS--GD}) one could pass from the GD to the
DS brackets.

Note also that one can rewrite the auxilary linear problem
as
\a
\frac{\pa \xi}{\pa x}(x)=Q(x) \xi (x)
\b
where $\xi$ is the transposed of an arbitrary row of
the matrix $V$ on $\F$ and
\a
Q(x)=\left (\begin{array}{ccccc}
0 & 1 & 0 & 0 & \cdots \\
-h & J & 1 & 0 & \cdots \\
0 & -2h & 2J & 1 & \cdots \\
0 & 0 & -3h & 3J & \ddots\\
\vdots & \vdots & \,&\ddots &\ddots \end{array}
\right )+d_0J(x)
\b
We thus conclude that $Q$ has  the same form
as the spectral matrix (the operator of multiplication by the spectral
parameter ) in the one--matrix models \cite{BX}. In contrast to the
CAT models, the first ("space") flow in the one--matrix models
is generated by the upper diagonal part
of $Q$, $Q_+$. This simple observation looks to be a promissing starting
point in relating the matrix models and the CAT.

We shall finish with the following remark.  Sinse CAT are
conformally invariant, they separate into two sectors
-- chiral and antichiral. In this section we considered the
conserved quantities in the chiral sector. The antichiral
currents are obtained through the same prosedure if
one starts from the Lax connection
\a
\L_+=\pa_++e^{2 ad \Phi}\E_1 ~~~~~~~~~~
\L_-=\pa_-+2\pa_-\Phi +\E_{-1}
\b

{\bf Acknowledgements }

It is a pleasure to thank  L. Bonora for several
interesting discussions,  suggestions  and for his kind hospitality
at SISSA. E. Aldrovandi is acknowledged for
his early collaboration and his interest in this work.
 I am also very grateful to E. Nissimov, S. Pacheva
and C. S. Xiong for discussions.
Partial financial support by INFN, Sez. di Trieste is gratefully
acknowledged.

\end{document}